# A Collaborative Intrusion Detection System Using Snort IDS Nodes

**Tom Davies[1], Max Hashem Eiza[2], Nathan Shone[2], Rob Lyon[2]**

[1]*Tom Davies, Radius Limited, Crewe CW1 6BD, UK.*
[2]*School of Computer Science and Mathematics, Liverpool John Moores University, Liverpool L3 3AF, UK.*

**Correspondence to:** Dr Max Hashem Eiza, School of Computer Science and Mathematics, Liverpool John Moores University, Liverpool L3 3AF, UK. E-mail: M.HashemEiza@ljmu.ac.uk; ORCID: https://orcid.org/0000-0001-9114-8577

## Abstract

Intrusion Detection Systems (IDSs) are integral to safeguarding networks by detecting and responding to threats from malicious traffic or compromised devices. However, standalone IDS deployments often fall short when addressing the increasing complexity and scale of modern cyberattacks. This paper proposes a Collaborative Intrusion Detection System (CIDS) that leverages Snort, an open-source network intrusion detection system, to enhance detection accuracy and reduce false positives. The proposed architecture connects multiple Snort IDS nodes to a centralised node and integrates with a Security Information and Event Management (SIEM) platform to facilitate real-time data sharing, correlation, and analysis. The CIDS design includes a scalable configuration of Snort sensors, a centralised database for log storage, and LogScale SIEM for advanced analytics and visualisation. By aggregating and analysing intrusion data from multiple nodes, the system enables improved detection of distributed and sophisticated attack patterns that standalone IDSs may miss. Performance evaluation against simulated attacks, including Nmap port scans and ICMP flood attacks, demonstrates our CIDS's ability to efficiently process large-scale network traffic, detect threats with higher accuracy, and reduce alert fatigue. This paper highlights the potential of CIDS in modern network environments and explores future enhancements, such as integrating machine learning for advanced threat detection and creating public datasets to support collaborative research. The proposed CIDS framework provides a promising foundation for building more resilient and adaptive network security systems.

## 1. INTRODUCTION

As the landscape of cybersecurity continues to change, security teams are facing an increasing number of challenges when it comes to identifying sophisticated cyber threats and malicious activities on computer systems. To monitor the network for any malicious activity, Intrusion Detection Systems (IDSs) are used. There are different types of IDSs which are used for different reasons depending on the type of the network [1]. When looking at basic IDSs, there are two main types characterised by their deployment and location: Host-based IDS (HIDS) and Network-based IDS (NIDS). HIDS monitors activities on a single host by analysing local system events/logs for any malicious behaviour (i.e., relies on anomaly detection). HIDS will determine if a system has been compromised and will alert administrators so they can act accordingly depending on rules that have been set up by the administrator of the IDS. Examples of HIDS are OSSEC [2] and Tripwire [3]. On the other hand, NIDS monitor network traffic and activities across multiple nodes (i.e., to investigate network traffic [4]). Typical examples of NIDS include Snort [5], Zeek [6], and Suricata [7]. Both types of IDS suffer from false positives and issues related to false alarm rate due to being overwhelmed by network traffic. When looking at the differences between HIDS and NIDS, generally speaking, HIDS rely on anomaly detection approaches to identify potentially malicious behaviour, while NIDS use signature detection [8]. Table 1 presents a breakdown of the key differences between HIDS and NIDS and provides examples of software that fall into these categories.

**Table 1. HIDS vs. NIDS**

| Feature | Host-based IDS (HIDS) | Network-based IDS (NIDS) |
|---|---|---|
| Focus | Monitors activity on a single host | Monitors network traffic on multiple hosts |
| Detection Scope | Protects individual hosts | Protects the entire network infrastructure |
| Visibility | Has visibility on host-based activity | Visibility into network activity. |
| Examples | OSSEC, Tripwire | Snort, Zeek, Suricata |

As cyber-attacks have become more advanced with respect to the tools and techniques used by adversaries, standalone IDS solutions face limitations preventing them from effectively detecting and responding to advanced cyber-attacks (e.g., DDoS attacks and Botnets). To alleviate these issues and improve the efficiency and availability of IDS, Collaborative Intrusion Detection Systems (CIDSs) are proposed. Note that CIDSs are also referred to as Collaborative Intrusion Detection Networks (CIDNs). In this paper, we use the term CIDS throughout for consistency except for some related works in the literature that specifically refer to CIDS as CIDN. CIDS is composed of cooperating IDSs that share knowledge to achieve better intrusion detection accuracy and reduce the number of missed/false alarms. This way, CIDSs have the potential to provide a more comprehensive defence against sophisticated cyber-attacks. However, research on CIDSs is still in progress to address issues such as the integrity of generated alerts in terms of correctness and completeness, and alerts' trustworthiness in relation to possible false alerts generated by compromised IDS nodes. Moreover, to the best of the authors' knowledge, there is no work that addresses these issues nor work that demonstrates a practical implementation to prove CIDS efficiency and accuracy against attacks.

In this paper, we are proposing a CIDS solution using Snort IDS nodes that demonstrates a lower malicious activities false positive rate and a higher detection rate in comparison to a single IDS node on a network. The proposed solution operates by sending the information generated from the Snort IDS at each individual node, to a central node, then on to a SIEM which will allow for correlation of data from several nodes to obtain a better understanding of the attacks faced by the network. This should allow for appropriate response from a security team. This should also allow for potential attacks to be identified more quickly compared to a single IDS, allowing for a faster incident response by a security team. The contributions of this paper are listed below.

- Conduct a thorough assessment of the CIDS setup using Snort IDS nodes to measure the effectiveness and accuracy of its intrusion detection capabilities. This contribution involves generating various simulated attacks and analysing the detection rates, false positive rates, and false negative rates of the system.
- Configuration and optimisation of Snort IDS nodes. This point focuses on the configuration and fine-tuning of Snort IDS to effectively detect and respond to cyber intrusions within the CIDS. It involves selecting appropriate detection rules, optimising performance, and minimising false positives/negatives to enhance the overall efficacy of the intrusion detection process.
- Analyse the collaborative decision-making within the CIDS by focusing on how Snort IDS instances communicate and share information. Evaluate the efficiency of information sharing, collaboration protocols, and the impact on overall detection accuracy. Assess the benefits and challenges associated with collaborative intrusion detection in a network environment.
- Finally, assess the security posture and response capabilities of the proposed CIDS to detect intrusions effectively. Examine the incident response mechanisms, logging and reporting capabilities, and integration with incident management systems. Measure the CIDS's ability to mitigate and respond to various types of attacks, including real-world scenarios, and assess the overall security posture achieved.

The rest of this paper is organised as follows. Section 2 describes the types of IDS and provides a background and domain analysis of CIDS including the state of the art. Section 3 describes the functional requirements of the proposed CIDS in detail. The design of the proposed CIDS (including the communication protocols, and Snort and SIEM configurations) is described in Section 4, where we also introduce different attack types to be used in the evaluation of the proposed CIDS. Section 5 outlines the technical details of developing and deploying the proposed CIDS in our testing environment. Testing and evaluation of the proposed CIDS are in Section 6. Finally, Section 7 presents our conclusions and sets out our future work.

## 2. STATE OF THE ART

As we mentioned above, an IDS is a device or software application that monitors a network and/or information system for malicious activities or policy violations. They respond to suspicious activities by warning the system administrator, displaying an alert, and logging the event. An IDS can be described as a function that classifies input data as either a normal event or an attack. It does so by indicating the absence (0) or presence (1) of an alert, represented mathematically as: IDS: $X \rightarrow \{0, 1\}$. An IDS may use signatures, anomaly-based techniques, or both. A signature-based IDS references a database of previous attack profiles and known system vulnerabilities to identify active intrusion attempts. Whereas a behaviour-based (or anomaly-based) IDS references a baseline or learned pattern of normal system activity to identify active intrusion attempts. A CIDS or CIDN is a system/network that connects IDSs to exchange information among them. Cooperation in IDSs enables the system to use collective information from other IDSs to provide more accurate intrusion detection locally or system wide.

The first paper that proposed combining multiple IDSs dates back to 2003 [9], where the authors examined the use of three elementary detectors placed at various system layers including Snort for network level, Libsafe for application level, and Sysmon for the kernel level. This approach provided a more detailed insight into of potential attacks on the network from varying viewpoints. The paper used various attack types, including Buffer overflow, Flooding attacks and Script-based attacks. By varying the attack types, authors could evaluate the CIDS against each attack type and its variants. These different attacks evaluated each detector and provided an improved overview of how successful these attacks have been and if they were detected or not.

In [10], the authors proposed a Snort based CIDS using blockchain and a Software Defined Network (SDN). The authors stated that their goal is to launch seven common attacks against the network and then look at the detection results. This paper proposed using an IDS built on a blockchain network to find the overall performance benefit against several attacks. The authors explained the design of their implementation - a Snort based IDS on a host-based and network-based design. They also described how using blockchain is of benefit against insider attacks. To evaluate their work, they carried out two different experiments in a modest testing environment that utilised three virtual machines, ultimately concluding that Snort IDS on a blockchain has a lower number of false positives. This conclusion is in agreement with other related works in the literature, suggesting that the future of networks using blockchain IDS is a very feasible way to solve current problems with IDS systems.

In [11], Fung focused on the threat of insider attacks against CIDNs. This research analysed the robustness of CIDNs against insider attacks which is a growing concern for organisations. The IDS considered in their work is a signature-based IDS where it explains how it will compare signatures with a trusted database in order to identify if an intrusion has taken place or not. This is an issue when looking at attackers targeting nodes on a network and providing them with signatures that can bypass the alarms of IDSs. Fung stated that many attacks against CIDN are able to bypass the system such as betrayal attacks and collusion attacks; this is what results in the false information being sent out to the nodes and is what compromises honest nodes. A robust design is needed to counter insider attacks to the CIDN. This shows that working towards a more secure and signature trust-based network using blockchain could benefit CIDNs and this can counter some of the issues relating to insider threats.

### 2.1. Collaborative Intrusion Detection System (CIDS)

The current strategy of CIDSs is split into several areas of research since, as mentioned before, traditional IDSs are falling behind more complex threats such as DDoS and Botnets. This situation makes a single IDS unable to analyse the vast amount of modern network data needing to be analysed, emphasising the need for a collaborative intrusion detection method. Wenjuan *et al.* discussed different ways to enhance the security of CIDNs against issues such as insider threats [12]. The main method is using challenge based CIDNs. A challenge based CIDN works as follows. Each network node chooses which other nodes it will associate with and therefore collaborate with. This allows for each node to understand the status of the other nodes and if they can be trusted. If a new node wishes to join the CIDN, it needs to verify its identity by getting a public-key pair from a reputable authority. However, such a system only becomes viable when there is a guarantee of trust between nodes, otherwise the system becomes vulnerable to insider threats which would render a CIDN ineffective.

When deploying a CIDS, there are three main architectures that can be used: centralised, decentralised, and distributed CIDS. A centralised CIDS consists of a single centralised analysis unit responsible for the collection of data, with several monitoring units to monitor the behaviour of their host or network [13]. The authors in [13] introduced the centralised CIDS approach to collaborative intrusion and discussed the issues arising from the centralised approach having a single point of failure. This is a common theme as single points of failure can be a target for a threat actor, and by targeting the central point of failure this would result in a loss of detection data. Nonetheless, the main benefit of a centralised CIDS architecture is that there is one central node that will do all the analysis from the other monitoring nodes, and this can lead to higher detection rates meaning lower false positives overall. The main disadvantage, however, is the issue of the logs going to one central analysis node which means this can be a target for an attacker. If this central node is disrupted, this can result in a loss of detections from the IDS. Figure 1 shows the three different types of CIDN architecture where M stands for Monitor node and A for Analysis node [13].

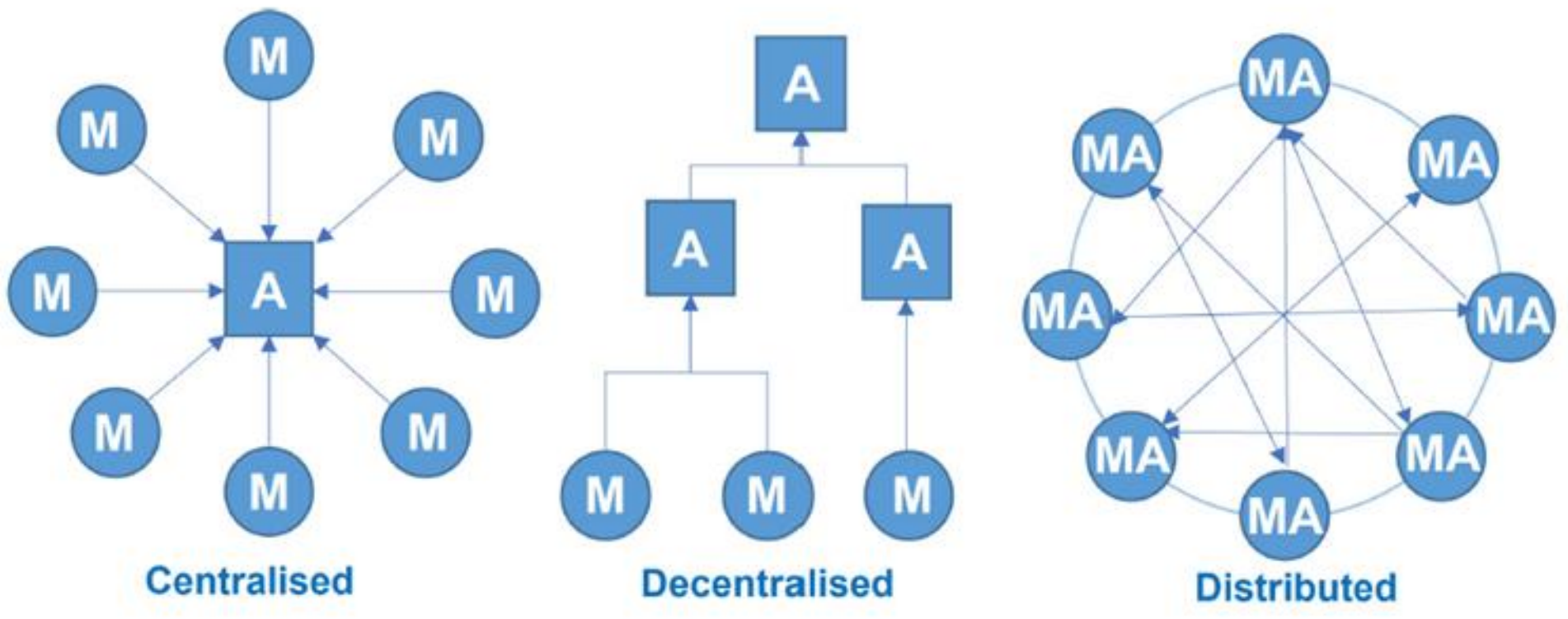

**Figure 1.** Overview of centralised, decentralised, and distributed CIDS architectures [13].

To summarise the advantages and disadvantages of each CIDS architecture including the hybridised centralised and distributed ones, we provide Table 2 below.

**Table 2. Advantages and disadvantages of each CIDN architecture**

| Feature | Centralised CIDS | Distributed CIDS | Hybrid CIDS |
|---|---|---|---|
| Deployment Model | Multiple IDS agents share information centrally | IDS agents distributed at network points | Combines centralised and distributed elements |
| Advantages | Improved detection accuracy through information sharing, better visibility across network | Scalability, improved performance, fault tolerance | Flexible deployment, leverages benefits of both |
| Disadvantages | Central point of failure, single point of attack, potential performance bottleneck | Complex configuration, management overhead, potential data security concerns | Increased complexity, requires careful design and integration |
| Suitability | Large, complex networks, require high detection accuracy | Decentralised networks, require high performance and scalability | Networks with diverse needs and security requirements |
| Scalability | Limited by central node capacity | Highly scalable | Depends on design and chosen components |
| Performance | Potential performance bottleneck at central node | Can offer high performance | Depends on design and chosen components |
| Cost | Higher cost due to central node and communication infrastructure | Lower cost for individual IDS agents | May vary depending on design complexity |
| Security | Single point of vulnerability, requires strong security measures | Potential data security risks in distributed environment | Needs rigorous security considerations for both centralized and distributed components |
| Management | Centralised management | Requires distributed management or central management tools | More complex management due to hybrid nature |

### *2.1.1 Trust Issues in CIDS*

One of the main issues raised in the CIDS literature is the issue of trust between the nodes, especially when considering the potential for insider threats on a network and ensuring the nodes have not become compromised. Therefore, establishing and maintaining trust amongst the participating nodes in the CIDS environment is the first challenge. Several solutions have been proposed to fix this problem, and these include the use of blockchain technology to ensure a secure and trusted decentralised environment can be created for the nodes to ensure that peer nodes have not been compromised [13]. Similar work in [14] considered the use of the blockchain to enhance the robustness of and effectiveness of signature based IDSs under adversarial scenarios (e.g., flooding and worm attacks) by sharing the signatures in a verifiable way.

In [15], Dawit *et al*. examined the suitability of using blockchain with CIDSs. They mentioned that using a blockchain network can stop attempted attacks due to the fact that if one block is altered on a node the other nodes will compare and therefore identify that the certain node has been tampered with. The authors discussed the actual application of blockchain can be used to overcome lots of the common problems single IDSs face such as Immutable logging, which can prevent IDS logs from being tampered. Blockchain technology can also be incorporated into the node itself to ensure no untrustworthy nodes join the CIDS. The paper explained about how the features of blockchain (e.g., transparency, integrity, immutability) mean that blockchain is perfect for CIDS since this can help trace back any active flow attacks. They also mentioned how the information is distributed along the nodes which is more secure than using cryptography alone. The paper presented a table demonstrating some of the key vulnerabilities of blockchain with examples of attacks that can be used to exploit these, this is very important for this paper, because it is one of the first papers to acknowledge that even though using blockchain in CIDS is good in theory, there are issues that need addressing if blockchain was to start being used on networks more widely.

## 3. REQUIREMENT ANALYSIS OF THE PROPOSED CIDS

For our proposed CIDS to function properly and solve the problem of single IDS failing against more complex attacks that a CIDS should be able to solve, the following components are needed:

- Snort IDS Sensors: These will be deployed across our testbed network, which is composed of 9 different network nodes. These sensors will be responsible for monitoring the traffic that is coming into the network and detect potential intrusions. There can be any number of Snort IDS sensors on the network. A total of 9 nodes have been used to provide a reasonable sample size that will allow for the comparison of a single IDS responding to attacks in comparison to several IDS nodes sharing information to have a better understanding of the attack scenarios.
- Central Node: This will be responsible for receiving the logs from the Snort sensors that will then be stored in a database, and this will then allow for the facilitation of communication between the logs from the sensors to the SIEM system.
- Database: This will be used to store the collected intrusion logs for historical analysis and reference in case

the SIEM has any issues. This way, a security analyst can still refer to the logs in the original format.

- SIEM system: This will be used to perform the advanced analysis and correlation of the intrusion data that has been collected from the snort sensors.

The proposed CIDS will require the following software and hardware to complete its implementation and evaluation later in this paper. These specifications will also allow researchers to duplicate our CIDS implementation and get similar results.

1. A server or PC capable of running several VMs at the same time including a VM running a SIEM log collector that requires many cores to run efficiently.
2. A virtualisation software that can handle different VLANs and allow for many snapshots to be captured for the development process.
3. Virtual machine ISOs for creating virtual machines, these include the following:
   a. Ubuntu 22.04 Desktop, for the hosts running snort IDS and for the central node.
   b. PFsense Firewall, used to deploy a software-based firewall for the network to facilitate correct VLANs for the LAN and WAN connections, this also ensures that only allowed traffic can go to the SIEM.
   c. Kali Linux, used for the attacking machine which will be preconfigured with all the necessary tools to perform testing against the proposed CIDS in the network.
4. Snort IDS version 3: This will be the IDS used to build the CIDS.
5. MySQL: This is the DB that will be used to store the data.
6. LogScale SIEM: This is the SIEM that will be used to correlate the data from the snort sensors.

Now that we have the proposed CIDS components, in the following we explain the functional requirements of each to ensure the CIDS can achieve its goals.

**Snort Sensors**

- Traffic Monitoring: Real-time network traffic monitoring is needed for Snort sensors.
- Intrusion Detection: Detect and classify potential intrusions based on predefined rules and signatures.
- Logging: Create thorough intrusion logs that include timestamps, source and destination IP addresses, and information about threats that have been discovered.
- Log Transmission: Send secure intrusion logs to the central node so they can be processed further.

**Central Node**

- Log Reception: Instantaneously obtain intrusion logs from Snort sensors.
- Authentication and Authorisation: Make sure that only approved sensors can submit logs by putting safe authentication procedures in place.
- Data Parsing: Extract pertinent information from incoming logs by parsing them.
- Database Insertion: Add parsed logs to the database so they may be retrieved later.
- Error Handling: Put in place procedures to deal with mistakes, like recording unsuccessful transmission attempts and alerting administrators.
- Communication with SIEM: Enable access to the stored intrusion data by facilitating integration with the SIEM system.

**Database**

- Schema Design: Create a database that will be able to effectively store Snort intrusion logs.
- Data Storage: Keep the logs secure by only having access to what is needed in the DB.
- Queries: Ensure the DB can be queried to obtain past intrusion data for historical investigations.
- Backup and Recovery: To avoid losing data in the event of system failure, the DB will be regularly backed up.

**SIEM System**

- Integration: The SIEM system must integrate with the central node to retrieve and analyse intrusion data.
- Correlation: The system must be able to correlate the intrusion events from the multiple nodes and be able to alert on patterns and potential security incidents.
- Alerting: Generate alerts for detected security threats these will relate to snort rules.
- Visualisation: Provide visual representation of intrusion data for easier analysis.

# 4. THE CIDS DESIGN

The design for our CIDS features a scalable architecture that can facilitate the collaborative effort from the Snort IDS nodes. The architecture is based on nine Ubuntu OS VMs all with Snort IDS deployed on them, which will then send the generated logs to a central node which has a MySQL DB installed. Then, the logs will be sent to LogScale, which will then be able to contextualise the data that the Snort nodes have generated. The reason for including LogScale in our design is because of the number of logs generated by Snort and the need for a quick understanding of the detections coming from Snort. If the logs are simply sent to the database, then this would become challenging for a security team to practically manage, preventing them from quickly and efficiently identifying the true nature of generated alerts.

Figure 2 illustrates a detailed network architecture diagram showing the nine network nodes that will send logs generated by Snort to the central node, which will then be inserted into the MySQL DB and then onto LogScale for data correlation for the security team to act on during incident response. The firewall provides a defence in depth approach to the network and will represent a more typical network with has multiple defensive technologies. In the following subsection, we discuss each component in Figure 2.

## 4.1. Snort IDS Nodes

These nodes are deployed across the network. Snort IDS nodes will monitor and analyse the network traffic for potential intrusions based on the ruleset provided to the IDS. This will generate logs containing the necessary information about detected events. The reason Snort was selected for this task as an open source and therefore low cost, yet industry standard IDS system. Snorts usefulness is also enhanced by community rules which can be used to help detect attacks against IDS and this also allows for a threat intelligence aspect to be built into Snort because as community rules are developed this will allow for new attacks to be identified by the IDS when they are discovered by the community.

In addition, Snort also uses signature-based detection to identify known attack patterns by matching network traffic against a set of predefined rules, which helps detect against known threats. Snort was selected as our IDS primarily due to its signature-based detection system. The enables known attack patterns to be matched against a set of well understood pre-defined rules, making the detection of known threats straightforward. It also allows us to identify attacks efficiently and intuitively in a collaborative way. The system is also known to be highly scalable, is open source, and comes equipped with a variety of community rules which help expand the IDS functionality.

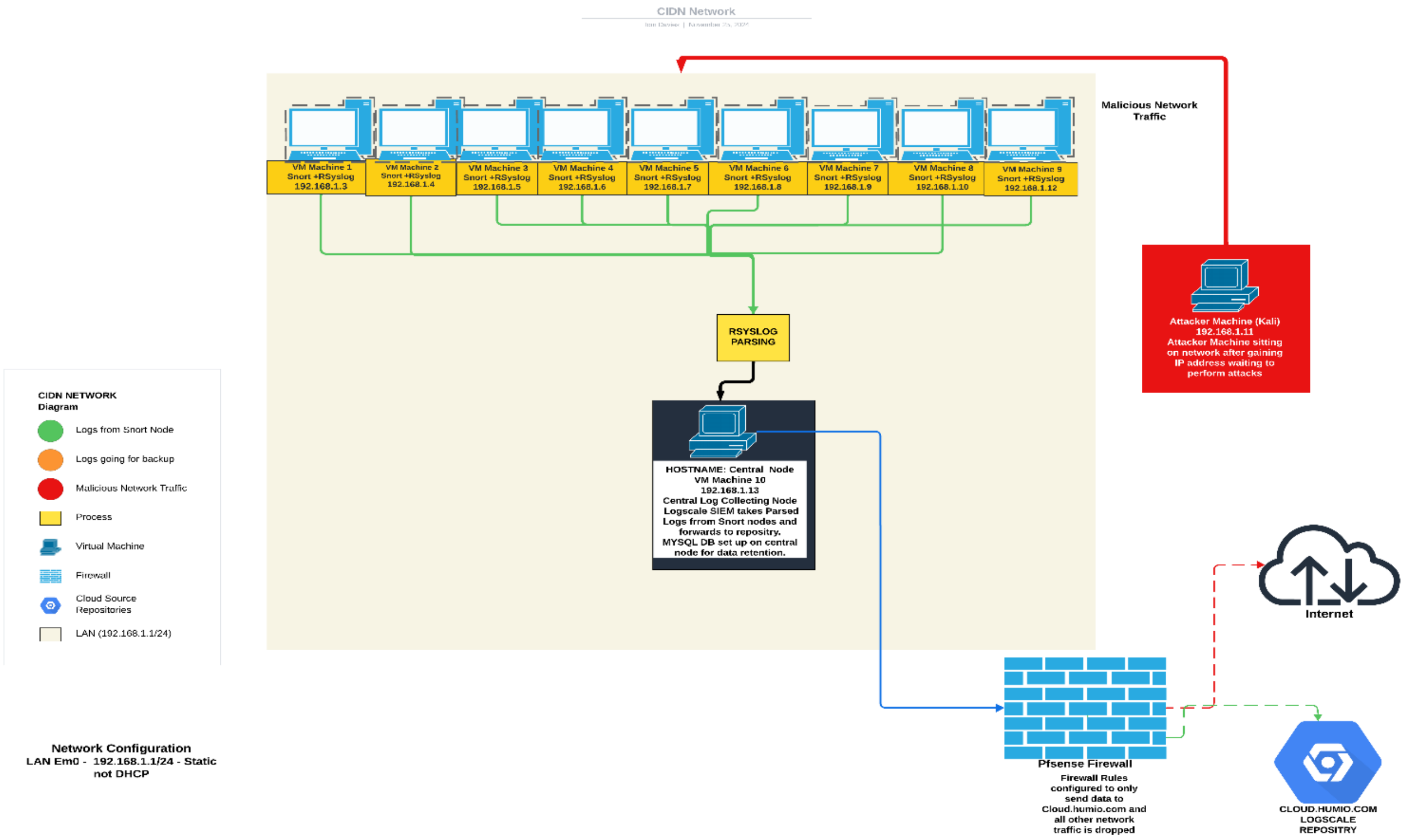

**Figure 2.** CIDS Design.

### *4.1.1 Snort Design Considerations*

This section will detail the Snort design we intend to use for our CIDS and the reasons behind choosing the configuration, along with how this benefits the design. The Snort nodes will all be running with identical configuration as there will be one virtual machine set up originally and then this will be used as the master VM image for all nodes, this will ensure that testing can be done for each of the tools used to set up the network successfully.

The first consideration is that we have opted to use Snort3. Snort3 supports hyper threading and allows shared memory, allowing for Snort to be more efficient at runtime [5]. These benefits mean that Snort3 will be able to scale much better than previous versions, this is important for two reasons, firstly the design of the CIDS needs to be scalable as this network design should allow for nodes to be added as the network grows without having any serious network performance issues. The second reason is that this version of Snort can process memory much more efficiently and therefore this will ensure that performance on each of the network nodes is not affecting the overall machine it is running on. The other reasons we have opted for Snort3 rather than Snort2 relates to the number of plugins that can be added to Snort, which allows for Snort to more modular IDS. It can also allow for future changes to be made to the IDS if a plugin was created in order to help benefit the CIDS. The last benefit of Snort3 is that this version has improvements such as supporting RegEx and sticky buffers, which also help with more flexible rule generation and futureproofing.

The second consideration for Snort will be the way it is running on the nodes. Snort has three modes of running, packet logging, sniffer, and IDS mode. We will be running Snort in IDS mode, but we will also have it running in promiscuous mode, this means that Snort will be capturing all the network traffic it sees unlike other network cards that will filter on the MAC address. This should allow for Snort to not miss packets that are travelling through the local area network which the nodes are on.

The third consideration is the how Snort will run on the nodes. We will run Snort as a system service so it will run automatically on startup. This is important because if a node was to go offline and restart without Snort running automatically then this IDS sensor will go down and we will not get the traffic from this. This also makes deployment easier as we will not need to spend time ensuring Snort runs correctly. The startup will call the `/usr/local/etc/snort/snort.lua` file which is Snort configuration. This configuration file will specify which interface snort will listen on, the interface will be configured by the firewall when creating the LAN and WAN.

### 4.2. Central Node

This will act as a collection hub to receive the logs from the nine network nodes. The central node will have a MySQL DB on it which will have a DB that will be where the logs are forwarded to from each node. The MySQL DB has two purposes in the CIDS design. Firstly, it is a central place for the logs to be stored and this will allow for logs to be checked if logs need referring to. Secondly, the purpose for having the logs sent to the DB is that the LogScale can then be used to collect the logs from the central node via a log collector that is deployed on the node.

### 4.3. SIEM System

Our CIDS uses the LogScale SIEM produced by CrowdStrike, formerly known as Humio [16]. The choice to integrate LogScale into our design is based on its ability to efficiently and effectively correlate data from diverse sources. SIEMs such as LogScale excel in data aggregation and analysis, enabling the identification of patterns and anomalies that are critical for intrusion detection. Additionally, the use of LogScale allows data from the network nodes to be visualised through customisable dashboards. These dashboards provide an intuitive and centralised view of network activity, simplifying the process of monitoring and understanding events within the network. Logs will be transmitted to the LogScale SIEM through a central log collector located on the Central Node. This configuration ensures that log data is sourced directly from the MySQL database, offering built-in data redundancy, and enhancing the reliability of the logging system. The other benefit of using LogScale SIEM compared to just using the MySQL DB for storing the logs, is that the SIEM allows for event correlation from up to 7 days from when the logs have been ingested, this will allow for a security team to look at the IDS logs from a longer period of time, this can help in the long term with the investigation of the alerts generated via the Snort IDS logs. The potential drawbacks to selecting LogScale are that with the community edition we are using for building our CIDS there is only 16GB ingestion a day, however this should not be an issue for our design as we only have 9 nodes and the logs generated by Snort are not that large. This issue of ingestion can also be fixed by the using the MySQL DB as a backup solution which will offload the issue of relying on the SIEM for keeping data for long periods which would incur costs.

The other SIEMs we have considered for our CIDS are Nagios [17] and Splunk [18]. Nagios does not provide some core features such as advanced querying, and the custom dashboards are not as detailed as those achievable with LogScale. The benefit of Nagios would have been that this SIEM also has a network analyser which could have worked in partnership with the IDS however this is only for the paid version and the free version of the SIEM is very limited. The other SIEM that we considered was Splunk [18]. Splunk is an industry leading SIEM and has a lot of features that would make the correlation of data much better and easier for an analyst to see the logs coming in from the central node. Yet there are no community editions of this SIEM, and you have to pay after a certain period of time, which would not work for the implementation phase in this paper. Table 3 provides a comparison between the three SIEMs we have considered in this paper.

**Table 3. Comparison of SIEMs**

| Feature | Logscale SIEM | Nagios | Splunk |
|---|---|---|---|
| **Log Collection** | Centralised log collection and aggregation | Centralised log collection | Centralised log collection and aggregation |
| **Security Analysis** | Advanced security analysis and correlation | Basic security analysis | Advanced security analysis and correlation |
| **Real-Time Alerts** | Real-time alerting for suspicious activities | Real-time alerts for system events | Real-time alerting for security incidents |
| **Scalability** | Scalable architecture for growing environments | Limited scalability options | Scalable architecture for large-scale deployments |
| **Integration** | Integrates with various security tools and systems | Limited integration capabilities | Extensive integration capabilities with third-party tools |
| **User Interface** | User-friendly interface for monitoring and analysis | Basic interface with text-based reporting | Intuitive interface with customisable dashboards |
| **Customisation** | Highly customisable for tailored security needs | Limited customisation options | Extensive customisation options for data analysis |

#### *4.3.1 SIEM Configuration Design*

To send logs to the SIEM, the flow of data moves from the nodes to the central node for the data to be stored within the MySQL DB, and LogScale will then collect the data from the MySQL DB. The reason for this configuration decision relates to the data retention needed for the logs, LogScale Community Edition only has a 7-day retention period used, therefore there needs to be data redundancy and backup with the DB. One benefit of having the logs stored in the DB is that even when the logs are shipped to the SIEM via the log collector, the logs can also be referred to in the DB, this would allow for a security team to use both the DB data and LogScale for investigating historic alerts. For the CIDS, LogScale is configured to accept syslog files, since the data generated via Snort will be outputted as syslog. LogScale will then be used to correlate the data using different queries within the platform. These queries can then be put within a dashboard that will automatically run and update as data is ingested from the central node, this will be able to give a real time understanding of the types of attacks that the Snort nodes are seeing and should allow for a security team to be able to respond to potential attacks more efficiently.

### 4.4. PFsense Firewall & Attacker Machine

Our CIDS design features a PFsense firewall to maintains overall control over the traffic traversing the network. The configuration used is to block all inbound and outbound WAN traffic except from the Central Node to LogScale. The attacker machine will be used to simulate attacks against the Snort nodes and will be on the same LAN. It is assumed an attacker has obtained physical access to the network, thus enabling them to launch attacks against the nodes in the network. This work will focus on simulating a small selection of network-based attacks, specifically DoS and port scanning. The reason for selecting these two, is that they are both attacks that can be directed at any host without requiring any initial exploitation and can easily demonstrate the benefits of a CIDS approach. Focusing on a small attack subset, enables evaluation of the implemented CIDS against real-world attack, whilst providing scope for further expansion in later works. This will test the collaborative approach to IDS due to if only some of the IDS nodes can identify this attack this can show the benefits of the IDS being collaborative compared to single ids nodes being deployed. Finally, we will launch an attack related to network recon such as DNS enumeration and reverse DNS lookups. This is important as the network design shows the attacker is already inside the network and therefore will want more details of the network environment to move across the network potentially laterally. This should allow us to detect, at the network level, a spike in non-existent domain responses due to the DNS enumeration.

### 4.5. Logging Protocols

The logging (i.e., communication) protocols that will be used to send the logs to the central node will be using TCP and the logs will be sent by configuring rsyslog to send the logs from the nodes to the central node. The reason for using rsyslog is due to the benefits this has in comparison to standard syslog. These benefits include the ability to call different modules relating to what you want rsyslog to do with the logs. The modules we will be calling for our network are ommysql module, this will be used to send the logs that are received from the nodes and this module allows for rsyslog to insert data into the MySQL DB. Table 4 provides breakdown of the main benefits of rsyslog in comparison to syslog.

**Table 4. Syslog vs rsyslog**

| Feature | Syslog | rsyslog |
|---|---|---|
| **Logging Protocol** | Standard logging protocol | Enhanced logging protocol |
| **Reliability** | Basic reliability | Improved reliability and robustness |
| **Performance** | Limited performance optimisation | Enhanced performance optimisation |
| **Configuration** | Limited configuration options | Extensive configuration options |
| **Filtering** | Basic filtering capabilities | Advanced filtering capabilities |
| **Forwarding** | Basic log forwarding capabilities | Enhanced log forwarding capabilities |
| **Compatibility** | Widely compatible with various systems and devices | Compatible with a wide range of systems and devices |
| **Logging Format** | Fixed logging format | Flexible logging format |
| **Centralized Logging** | Supports centralised logging with limitations | Better support for centralised logging |

## 5. DEVELOPMENT AND DEPLOYMENT OF THE PROPOSED CIDS

### 5.1. Snort IDS Configuration and Setup

To run Snort IDS on each VM, the following configuration needs to be done: Snort Config File / logging and output, Networking interfaces, and Rule management. First, we start with Snort3 configuration file called snort.lua, which contains information such as specifying what logging snort will do. The first change in the configuration file is the output that Snort will output to, the chosen output will be syslog as this will then be shipped to the central node via rsyslog. In order to specify the syslog output, the configuration will be as follows in Figure 3.

```
-alert_syslog Syslog configuration --
alert_syslog =
{
   facilty = local7, -- auth: alert_syslog.facility | part of priority applied to each message
   level = alert, -- info: alert_syslog.level |
   options = pid -- options | alert_syslog.options | Used to open the syslog connection
}
- End of syslog configuration --
```

**Figure 3.** Syslog configuration in snort.lua

It calls the Snort configuration file which will specify that Snort will output and alert in a syslog format and the logs will be stored in `/var/log/snort`, this will allow for rsyslog to ship the logs generated from the log file. Snort will run in promiscuous mode, which will allow for it to look at all traffic on the interface and better make decisions when it relates to looking at the rules and then alerting accordingly.

### *5.1.1 Snort Rules*

The rules in Snort are responsible for generating the alerts when traffic is detected that matches a signature or attack pattern. The rules in Snort3 are different from previous versions due to the fact that they are more standardised and allow for more flexibility when it comes to writing the rules. Snort3 also allows for the rules to be tuned to be more specific, which means that the detection accuracy should be increased. There are many community rules that could be used for this Snort deployment; however, we have opted for using just local rules that we wrote relating to the specific attack scenarios, which we will outline in the testing phase of this paper.

```
sysadmin@node: /etc/rsyslog.d                    sysadmin@node: /usr/local/etc/rules
 GNU nano 6.2                                    local.rules *
alert icmp any any -> any any ( msg:"ICMP Traffic Detected"; sid:10000001; metadata:policy security-ips alert; )
alert tcp any any -> any any (msg:"NMAP TCP SCAN"; sid:10000005;)
```

**Figure 4.** Example of Snort rules in Snort3

This shows that there are multiple additions to a basic rule you can add, the rules in Figure 4 will not be the final rules that will be used but these have been used to validate that Snort is running correctly and logging to the correct area. The rules also allow for the rule to trigger against policies if the user wants. This means that Snort can be expanded to provide a much more efficient and tailored alerting mechanism against attacks. The logs that are generate by these rules are then shipped using rsyslog, which will be covered in sub-section 5.3.

## 5.2. Firewall Configuration and Setup

The network design has a firewall, the reason for this is to provide a defence in depth approach. The firewall will allow for the network to be segmented into a LAN which all the machines are on, the LAN will then have firewall rules which will mean that only traffic is allowed to go to the SIEM repository which is “cloud.community.humio.com” while any other traffic will be dropped. This ensures the only external connection coming from the LAN will be to LogScale. To set up the firewall, first the VM needs to run through the setup process, this will allow for the interfaces to be specified, these interfaces are as follows: `LAN (lan) --> em1  --> v4: 192.168.1.1/24`. These interfaces are used to have a WAN and LAN in order for firewall rules to be applied to block all incoming traffic to the LAN and to the WAN. The IP range is 192.168.1.x for the LAN, the VMs will then join this LAN via the LAN segment which will then allow them to have a static IP on the LAN network. This will have the nine nodes and the central node on this network.
Using PFsense also allows for logging to ensure that if any malicious traffic attempted to probe the network externally it would block and flag it. The firewall logs also allow us to monitor the traffic and ensure that the data is going to LogScale as it is meant to. Finally, the firewall rules for the setup will have two rules, one to block all the traffic exiting the LAN and then another one which only allows traffic to LogScale, this is based on the URL and IP range for the SIEM repository as shown in Figure 5 below.

Rules (Drag to Change Order)

| | States | Protocol | Source | Port | Destination | Port | Gateway | Queue | Schedule | Description | Actions |
|---|---|---|---|---|---|---|---|---|---|---|---|
| | 1/1.56 MiB | * | * | * | LAN Address | 443 80 | * | * | | Anti-Lockout Rule | |
| | 1/4.18 MiB | IPv4 TCP/UDP | * | * | Humio | * | * | none | | This is a rule to only allow traffic to humio | |
| | 0/2.41 MiB | IPv4 TCP | * | * | * | * | * | none | | Blocked traffic (allow default) | |

**Figure 5.** Firewall Rules

## 5.3. SIEM Configuration and Setup

As mentioned before, the SIEM will be responsible for aggregating and querying the data received from the nine nodes with the logs generated via the IDS nodes. First, we need to deploy the log collector onto the central node where all the logs will be shipped to from the IDS nodes. Once installed from LogScale, there are a few options in relation to how the SIEM log collector can be configured to run. The options are automatically or manually start/run/stop the log collector. The benefits and drawbacks to these two approaches are that if the log collector is running constantly this creates a higher performance overhead because of the log ingestion from the Central Node to the SIEM. However, because of the nature of the CIDS and the constant running of the IDS nodes shipping logs to the central node, we need the Log collector to be constantly running.

### *5.3.1 Enrolment of A Node*

To ensure the logs can be sent to the SIEM, in the SIEM repository setup, the collector instances need to be set up and have a config specified. The config will be used to specify what logs the log collector should collect, in this case it will be syslog as this is the log type that is coming from the IDS nodes. Once a configuration is specified, we then enrol the host by using enrol and then specifying the specific enrolment token that has been generated as shown in Figure 6 below. Once the enrolment is successful, the node will then show in fleet overview along with some statistics such as ingestion and CPU usage. This is useful, because this will allow for the node to be monitored, and this should help point out any issues during the log ingestion into the SIEM.

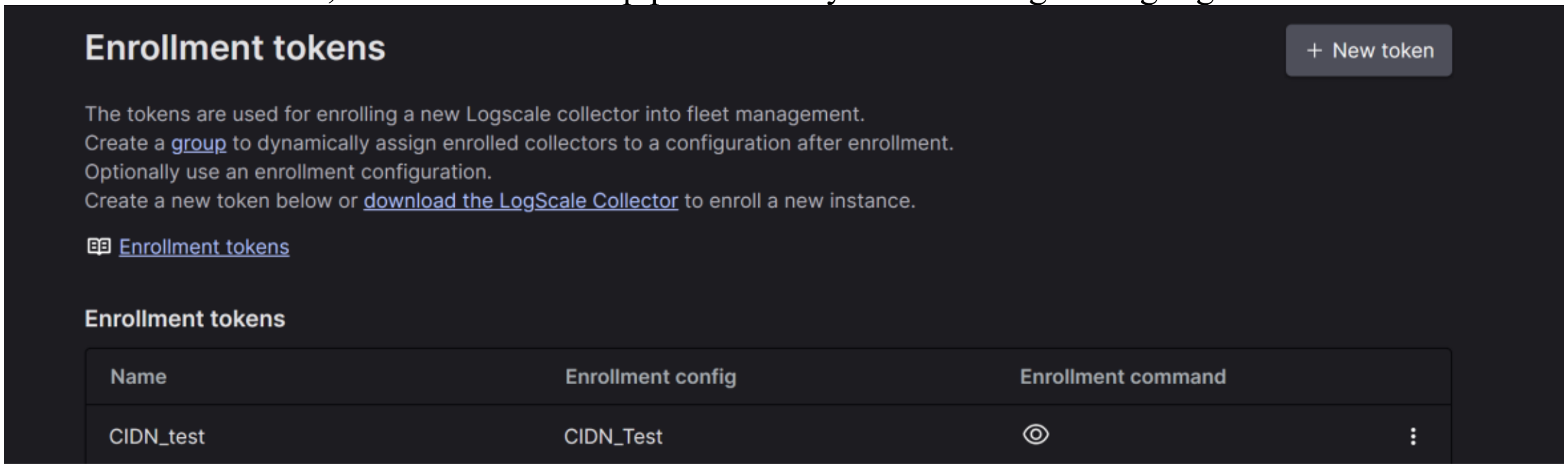

**Figure 6.** Enrolment page which generates the enrolment token

### *5.3.2 Configuration for Syslog collection*

Once the node is enrolled, the focus for finishing the SIEM configuration relates to the configuration for the logs that the log collector will pull. The nodes then ship the logs using rsyslog to the central node on port 514 therefore the SIEM needs to collect logs as the Central node receives them as well as inputting into the DB for backup purposes.

```
 GNU nano 6.2                                                              33-logscale.conf
odule(load="omelasticsearch")
emplate(name="logscaletemplate" type="list" option.json="on") {
 constant(value="{")
    constant(value="\"@timestamp\":\"") property(name="timereported" dateFormat="rfc3339")
    constant(value="\",\"message\":\"") property(name="msg")
    constant(value="\",\"host\":\"") property(name="hostname")
    constant(value="\",\"severity\":\"") property(name="syslogseverity-text")
    constant(value="\",\"facility\":\"") property(name="syslogfacility-text")
    constant(value="\",\"syslogtag\":\"") property(name="syslogtag")
    constant(value="\",\"name\":\"") property(name="programname")
    constant(value="\",\"pid\":\"") property(name="procid")
    constant(value="\"}")

.* action(type="omelasticsearch"
   server="https://cloud.community.humio.com:443"
   template="logscaletemplate"
   uid="snort" pwd=[redacted]
   bulkmode="on"
   usehttps="on")
```

**Figure 7.** Configuration for SIEM to collect syslog on Central node (token is redacted for privacy)

The configuration, in Figure 7, uses the rsyslog module `omelasticsearch` which is used as the parser for sending logs to the SIEM. This rsyslog configuration is created in the rsyslog.d folder and the configuration file for LogScale is created called `33-logscale.conf`. The module is then loaded by rsyslog and the data is set into a template so that LogScale can accept it into the SIEM. The *uid* and *pwd* field refer to the ingestion credentials generated within LogScale to allow data to be sent to the CIDS repository. The next step is to check that the data is being sent to the SIEM, this is done by checking the SIEM in the repository and doing the following query `Count(syslogtag)` that checks that the data with the tag syslog is being sent to the SIEM.

## 5.4. Rsyslog Configuration and Setup

### *5.4.1 IDS Node Rsyslog Configuration*

The communication protocol that will be used to ship the logs from IDS nodes to the Central node will be Rsyslog which is a more advanced version of syslog. The first stage is to ensure that rsyslog is running on each node. The configuration of rsyslog needs to be that it ships the logs generated by Snort to the central node which is on the same LAN. The configuration for rsyslog is broken down into two configuration files, one is `rsyslog.conf` and the other is `rsyslog.d` which is where you can have multiple configuration files. The first configuration file relates to different modules that rsyslog can use in order to log data, the two modules we will load for the configuration on the IDS nodes are as follows.

```
# provides UDP syslog reception
module(load="imudp")
input(type="imudp" port="514")
# provides TCP syslog reception
module(load="imtcp")
input(type="imtcp" port="514")
```

The modules for rsyslog will be responsible for taking any data that has been specified and then send it via tcp and udp on port 514. The final section of the configuration file specifies where rsyslog should go in order to collect the logs from.

```
# Include all config files in /etc/rsyslog.d/
#
$IncludeConfig /etc/rsyslog.d/*.conf

#template to store syslog messages
$template RemInputLogs, "/var/log/snort/snort.pid/"
*.* ?RemInputLogs
```

The rsyslog configuration will specify the log sources, this is in Snort logs folder which is where the Snort logs that are generated will be stored in. The logs will be collected and forwarded to the central node by specifying the forwarding IP which is found in `rsyslog.d` config file. In `rsyslog.d` config file, this relates to any rules that rsyslog must follow, this statement tells rsyslog to send the logs it received to send it to the following IP which is the central node: `*.* @192.168.1.13`. The configuration is now working for nodes 1-9. The next step is to set up rsyslog on the receiving server (central node) in order to receive the logs successfully.

### *5.4.2 Central Node Rsyslog Configuration*

The central node will have more rsyslog modules loaded due to it needing to input the received logs into the Database for the data retention for the CIDS. The first configuration needed will be the same as the nodes as the TCP and UDP syslog needs to be enabled for rsyslog to open port 514 to accept data.

```
# provides UDP syslog reception
module(load="imudp")
input(type="imudp" port="514")
# provides TCP syslog reception
module(load="imtcp")
input(type="imtcp" port="514")
```

For rsyslog to be able to put the logs into the database, it needs to call a MySQL module which will be used to insert data sent from syslog into the specified Database. The configuration for the database will be discussed in subsection 5.5, the rsyslog configuration is as follows.

```
module (load="ommysql")
*.* action(type="ommysql" server="localhost" db="snort" uid="snort" pwd="test")
```

Once the configuration has been done, we can verify that the configuration is not failing by checking the status of rsyslog, this is showing as running on both the nodes and the central node therefore there is no issues relating to logs being sent or received or inputted into the database.

### 5.5. Database Configuration and Setup

The DB will be used to store the logs from the IDS nodes. This will act as a backup solution for the logs but also as a reference point in the event the SIEM is unavailable due to any issues. The DB will not have the same aggregating and querying capabilities as the SIEM, but the DB will be able to show an investigator what traffic is coming from the IDS nodes and any alerts that are generated. The choice of DB will be MySQL, which is installed when on the Central node. The DB has a root user however we will create a user for rsyslog to be able to access the DB, this will ensure that rsyslog can input data however it cannot do anything apart from insert data. This is best practice when creating an automated system for inputting data. We will be using DBBeaver [19] to visualise the data and see the logs in a more effective way instead of the terminal.

The first step to configure the DB is to create the DB itself which is called snort, this is done by the using the command CREATE DATABASE snort; Then, we create a table called SystemEvents which will be the table that will have the log data sent from the nodes in the DB. The system events table will have the following fields as shown in the Entity Relationship Diagram (ERD) in Figure 8 below. Note that the message column has a long text datatype because it will be able to account for the different outputs from the different Snort rules that will be used as this is what will be in the message section.

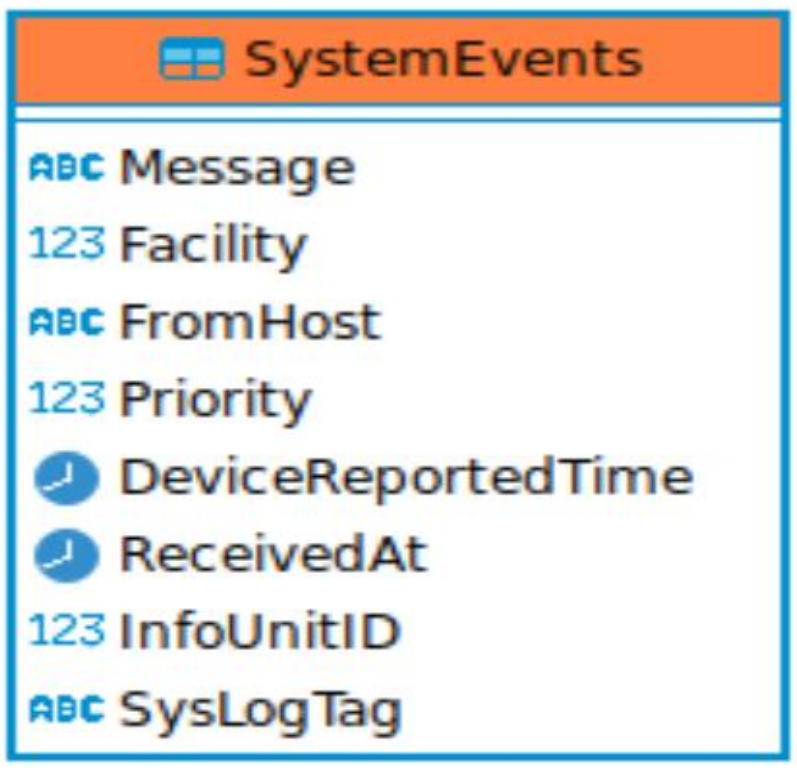

**Figure 8.** ERD diagram showing fields in SystemEvents table

Figure 9 shows an example of how rsyslog is successfully able to input different Snort logs from different nodes, this also shows that the DB has the ability to handle different alerts that snort could generate because of the long text field in the DB.

| | | | | |
|---|---|---|---|---|
| [1:10000001:0] "ICMP Traffic Detected" {ICMP} 192.16ɛ | 4 | node | 6 | 2024-0: |
| [1:10000001:0] "ICMP Traffic Detected" {ICMP} 192.16ɛ | 4 | node | 6 | 2024-0: |
| [1:10000001:0] "ICMP Traffic Detected" {ICMP} 192.16ɛ | 4 | node | 6 | 2024-0: |
| [122:23:1] "(port_scan) UDP filtered portsweep" [Priority | 4 | node4 | 6 | 2024-0: |

**Figure 9.** DB example of data inserted from Snort logs

## 6. TESTING AND EVALUATION OF THE DEVELOPED CIDS

The testing scenarios that will be used to test the detection capabilities of the proposed CIDS will relate to the attack types that were mentioned in Section 4.4. These attacks will come from the attacker machine and will first attack a single IDS node and using the SIEM, the data will then be analysed to see how many alerts are generated in comparison to the attack and if any of the attacks have been missed due to the IDS missing the attack type. The second testing phase will relate to attacking several of the IDS nodes with the same attack type. This will then allow for the SIEM to correlate the data from multiple IDS nodes and to be able to determine whether the increase in IDS nodes on the network sending data to a centralised point will allow for a better understanding of the attack that is going on in the network. This will be able to see if there is a pattern of attack, for example is a certain node being targeted more than others. This will also be able to test how well the CIDS scales in relation to the data ingestion into the SIEM and the data that is also going into the DB. The tests will allow for a better understanding of any drawbacks to the implementation of the CIDS that has been created.
The tools that will be used to perform these attacks will be tools that can be found in the default build of Kali Linux, the port scans will be done by Nmap and the Ping flood attacks will be done using a tool called hping3, this tool is able to flood a victim with either SYN or ICMP requests which means that the host will be overwhelmed with network requests. This means that the attacker can perform other attacks against the host which the IDS may miss because it will be monitoring a network that is being flooded with ping attacks by the attacker. These test cases will relate to network attacks an attacker can do when they are inside the network to either gather more information on the network or to disrupt the network with a DoS type attack. There will be four test cases for this test. The first case will test a single node and then the second case will test a variety of nodes on the network to see how the IDS react and then by using the SIEM the results will be able to be aggregated together to determine if a single IDS has a higher or lower False Positive (FP) rate. Also, when testing multiple nodes, it will also be able to test how the network handles a lot of traffic going to a centralised location. This will be able to test whether the CIDS can scale and any issues that are found during testing can be evaluated. A summary of test cases and their objectives is given in Table 5 below.

**Table 5. Testing table breaking down attack types and which nodes will be attacked**

| Test case ID | Attack Type being tested | Nodes being attacked | Snort rule being tested | Objective testing relates to |
|---|---|---|---|---|
| TC-1 | Nmap TCP port Scan | Node1 | Nmap Port scan rule | Testing FP and TP rate |
| TC-2 | Nmap TCP port Scan | Node2, Node4, Node6 | Nmap Port scan rule | Testing FP and TP rate and Scalability of CIDN |
| TC -3 | Ping Flood Attack | Node1 | Ping Flood attack rule | Testing FP and TP rate |
| TC -4 | Ping Flood Attack | Node3, Node5, Node7, Node9 | Ping flood attack | Testing FP and TP rate and Scalability of CIDN |

## 6.1. Snort Rules for Detecting Attacks

This is a breakdown of the Snort rules that will be used to detect the different attacks performed against the IDS nodes. The reason for creating the custom Snort rules instead of using Snort community rules is because of the specifics of the types of attacks that have been decided for this particular testing case. However, any of Snort rules will work with being shipped from the nodes to the central node and to the DB and SIEM. This is because the logs are parsed so that the SIEM can accept any logs generated by Snort as they are created in the format specified in the .lua file.

**Detection of potential flood attack**

```
alert icmp any any -> any any (msg:"ICMP Flood Detected"; detection_filter: track by_dst, count 150, seconds 3; classtype:bad-unknown; sid:100001; rev;1;)
```

This rule monitors the rate of ICMP packets directed at the destination IP. If the number of packets is exceeded by the threshold (this has been set to 150 packets within 3 seconds) this rule will trigger and alert against a potential ICMP flood attack.

**Detection for potential Nmap Scan**

```
alert tcp any any ->any any (msg: "NMAP TCP Scan";sid:10000005; rev:2; )
```

This rule monitors for any TCP traffic that could potentially be nmap scanning the network, this can be made more specific to monitor traffic on specific ports, but this rule is keeping it general currently.

## 6.1. Port Scan Testing

The first test **TC1** is for testing how the IDS deals with port scanning against a single IDS. The rule in Snort should detect the attempted scan and alert that this node is being scanned and should display the attackers machine IP. When scanning one IDS node, the Snort rule alerts and generates a lot of alerts because of how broad the rule is in relation to the IP. This results in a lot of messages coming into the SIEM when node1 is being scanned. When changing the rule to just look at traffic coming into the host being tested the following results are shown.

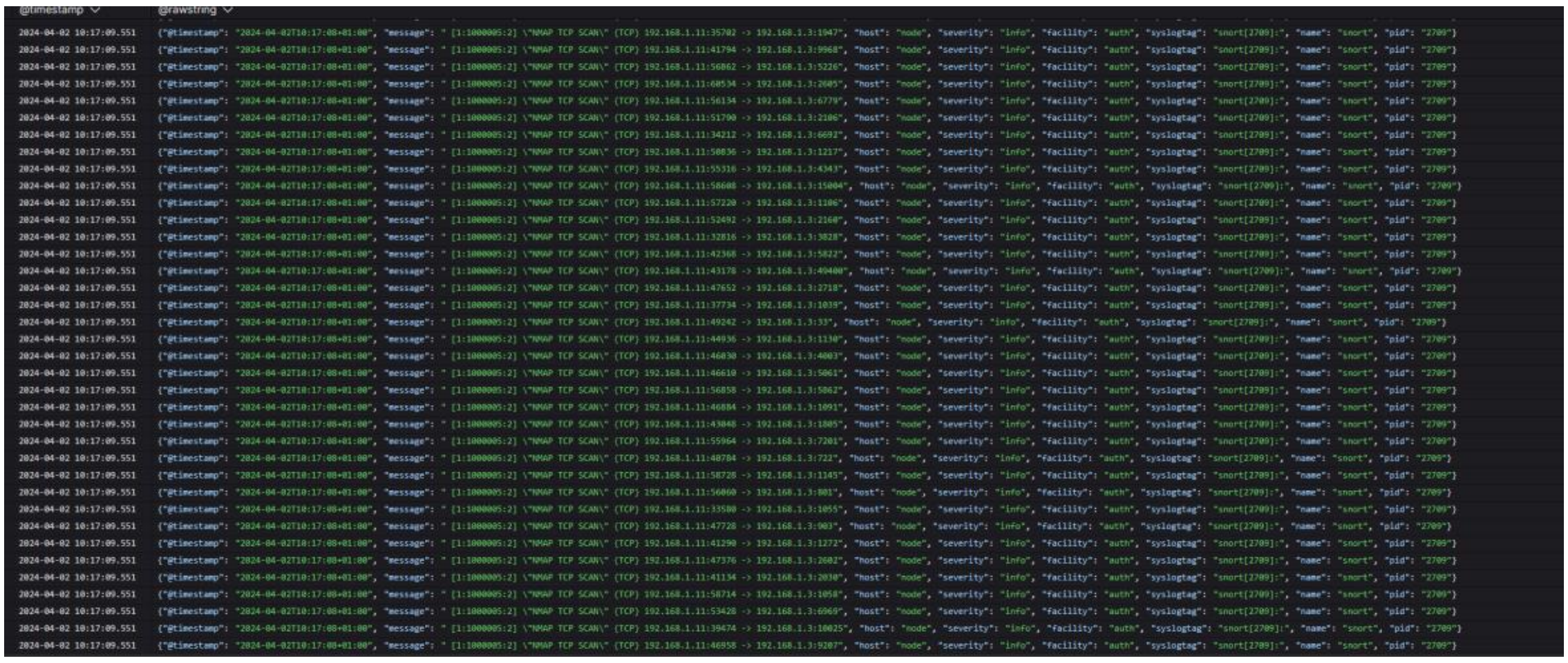

**Figure 10.** Event stream showing that an NMAP scan is happening against the node

When using `count()` in the SIEM, the result shows that the rule is working against the nmap scan and due to the amount of traffic that is generated by an nmap scan, the rule generates a separate message for every port that is scanned therefore giving the total count as 656 in this example.

This gives a baseline for how a single IDS responds to an nmap scan. The next test case will be scanning multiple nodes to test how the CIDS design handles more data being ingested and to see if the CIDS can give more accurate results in which shows that the SIEM has correctly ingested the data from the nodes in relation to nmap scan. This also shows that the nmap rule may need to be tuned so it does not generate an alert for every port but one single alert since it is only a single nmap scan being performed and the number of events ingested can show that there are multiple nmap scans happening when in fact it is only one. This would generate a lot of false positives due to the number of events for a single nmap scan when it should just generate one single detection so that there is not alert fatigue. This can be fixed with changing the rule threshold for detecting nmap scans.

When testing multiple nodes with the same nmap scan rule (i.e., **TC2**), the first analysis that will be done in the SIEM is to see the number of messages coming into the SIEM during the same period of the attack in each test case. When querying the SIEM using `count(syslogtag)` this will show the number of messages that come into the SIEM during a time interval, which is 2837 in this case. This shows that there are more than the three times of expected events as there was for the single IDS node when tested during the same time frame and under the same attack. This can give a security team a better understanding that a wider attack is going on within the network and will allow for further investigation. This also looks to have the similar issue to the number of events relating to the nmap scan, this is due to the rule generating an alert against each port scanned and not an individual alert relating to one overall nmap scan being ran against the node. Figure 11 below shows the increase in traffic from snort during testing the nmap rule on multiple nodes. This test case has shown that when there are multiple nodes being scanned and sending data to the SIEM, the SIEM is able to count the number of messages coming into the SIEM for comparison with single IDS node traffic. There are also no issues relating to the data being ingested into the SIEM which means that a security team can then further analyse and query the data within the SIEM to understand what is happening on the network and which nodes are being attacked.

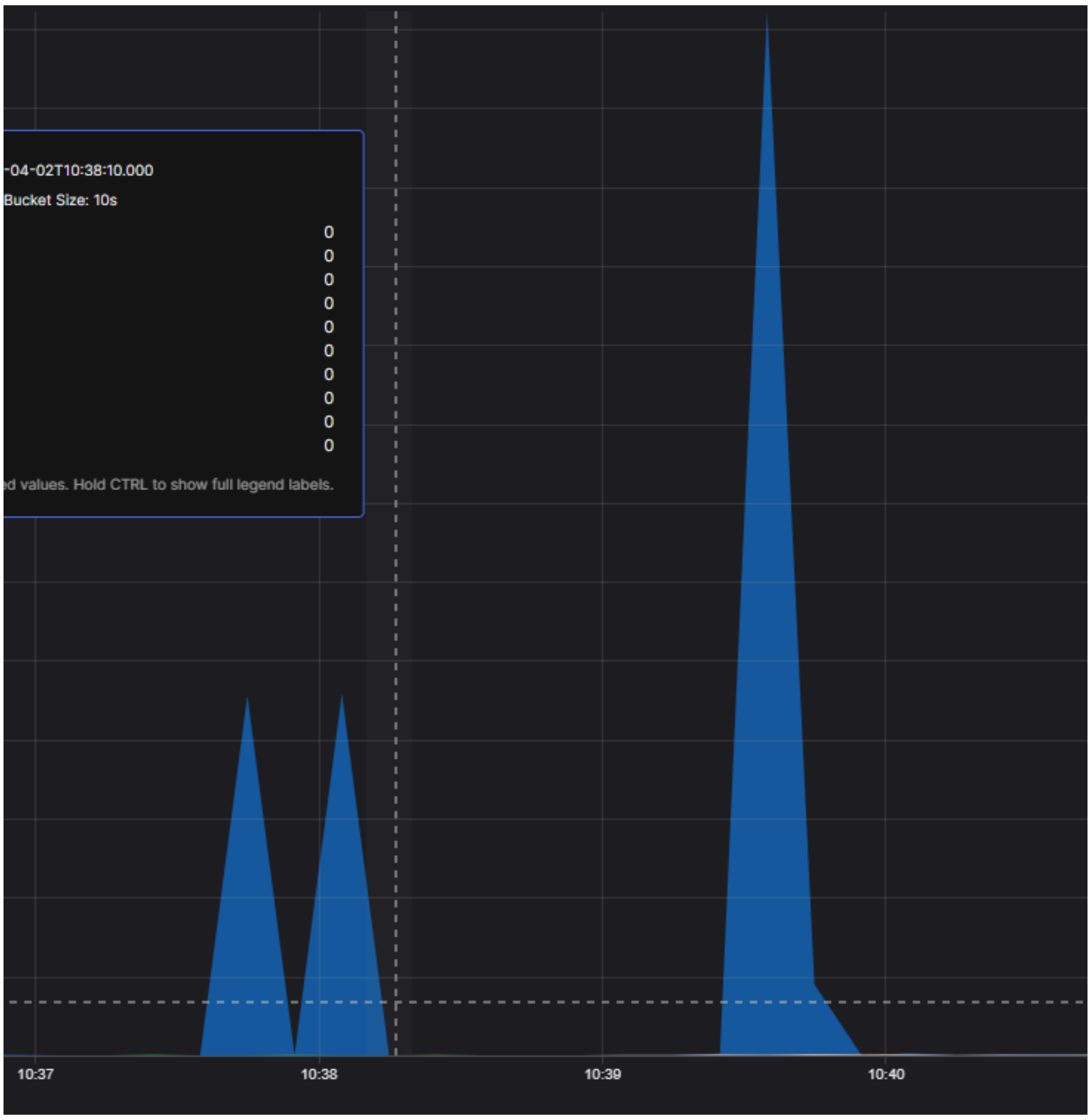

**Figure 11.** Time frame showing the sudden increase in Snort logs during the testing

## 6.2. Ping Flood Testing

The next test case will check how the CIDS responds under a significant amount of traffic from an ICMP flood attack, which is **TC3**. The tool that will be used to perform the ping flood attack will be hping3. The syntax of the command is `hping3 -1` which specifies that we will use the icmp traffic and the `--flood` command is used to flood the target with ICMP packets. The `-V` flag is used to verify that the flood attack has started successfully. Looking at the SIEM events and aggregating how many messages with the syslog tag are sent during the time frame, we can see there is over 56,000 events. We can use the filter in the SIEM to see the traffic coming from the host that is being attacked. This event stream in Figure 12 shows that there is a significant amount of data coming from the node during this attack. The results in this event stream are all TP because of how the rule has been configured that there must be a significant rate of ICMP packets for the rule to trigger. The issue with the rule is that it will constantly generate until the ping attack stops. However, the events can be filtered down so the security team can identify when the attack started, the message also is able to identify that the attack is coming from inside the network therefore meaning that this could be blocked potentially when it alerts the SIEM.

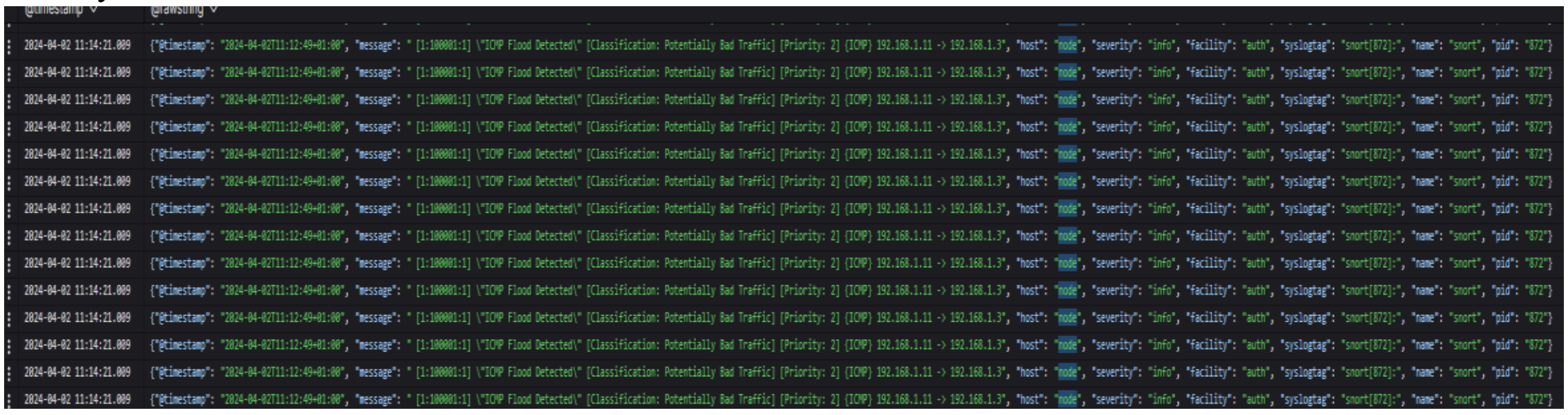

**Figure 12.** Event stream showing the ICMP flood in event stream for node

This test case shows that the current configuration for this rule generates a lot of alerts when the ping attack is ongoing, this could cause an issue for the security teams dealing with the detection due to the number of alerts being generated. The alerts generated by a single IDS would need tuning so that the security team looking at the data in the SIEM does not have alert fatigue due to the huge number of alerts. The number of alerts being over 56,000 events are a lot for the single IDS node however this is due to the amount of ICMP traffic the node is experiencing. The TP of this attack type will be high due to the amount of traffic but normal ping requests during this time frame will also be shown in the event stream in the SIEM due to once the threshold has been met for the rule it will detect. This can be fixed with a higher threshold for detecting ICMP attacks for the individual nodes.

The next test case **TC4** will test the issues that a DoS attack can have on the CIDS and does it affect the response capability of the IDS. The single IDS generated a lot of events during a one-minute time frame of a ICMP DoS however with multiple nodes being attacked at the same time the network may not be able to send all messages to the SIEM. The same attack is done against the IDS nodes 3,5,7,9. The results are as follows. The number of alters was over 103,000 alerts which is expected since four IDS nodes are under attack. The FP of the data in this time frame looks to be lower than a single IDS because with the amount of data coming from each node there would be the expectation that there would be 4 times the amount of data being ingested into the SIEM based on the results from the single IDS node being attacked. The data ingested can show that multiple nodes are being attacked at the same time with the same ping attack, filtering in the SIEM will allow for the security team to see that this is coming from the same IP therefore allowing for remediation action to take place.

There also looks to be a network issue for how fast rsyslog can process the number of logs, this could be fixed by adding rate limiting to the configuration of rsyslog. Even with the rate limiting from rsyslog it looks that the IDS can handle the extreme number of packets it has to process and check against the rules. There seems to be logs from every node that has been attacked which means that there is no issue relating to any of the IDS failing during this DoS attack however a more sustained attack may be able to stop the nodes from sending data to the SIEM. This test also shows that the network can respond to an extreme amount of data passing through it to the centralised node which is successfully sending data to the SIEM and DB. This shows that when multiple nodes have a lot of data being sent from each node to the centralised node, the node can handle the data coming into the node and is able to process it effectively and send the data to a DB and onto the SIEM without a significant slowdown in relation to the time from attack to the alerts in the SIEM.

### 6.3. Overall Evaluation of the Results

The developed CIDS has been built and proved to be able to successfully collect data from the several nodes on the network under different network attacks. It demonstrated that by using the SIEM, we are able to better understand what is happening in the network and then by drilling down into the data sent from Snort, a security team should be able to respond effectively to attacks on the network. The IDS configuration showed that it can effectively still detect even during an increase in network traffic from a DOS attack. This is due to the configuration of the Snort IDS on the nodes which means that it is very efficient and would take a lot of traffic to make the IDS stop working effectively. When testing the scalability, the events sent to the SIEM appear to be ingesting correctly however when testing the nodes with the ICMP attack it was found that some of the rsyslog configuration means that the number of events arriving in the SIEM are not as many as would be expected compared to a single IDS. However, this can be fixed by implementing the rate limiting within the rsyslog configuration design however this could be fixed with time for tuning and retesting the CIDN against the same attacks.

The developed CIDS has several improvements that can be made to make it more efficient. In relation to the number of events that are sent to the SIEM, this can be done by smarter rules and better configuration relating to how rsyslog ships the data to the SIEM. In terms of testing, the tests that have been done in this paper are only network-based attacks however the CIDS design means it would be capable of alerting against malware using Snort's community rules which would be interesting to test to see how the events are shipped due to the way some community rules work in relation to the fields it would generate. Other test cases that could have been tested relate to malware and testing to see how the CIDS reacts to prebuilt community rules relating to malware signatures, this would move away from network attacks and more targeted host-based attacks in which the IDS can try to detect against malware on the compromised hosts.

Overall, the developed CIDS demonstrates the principles of how collaborative IDS is able to better help security teams identify attacks on the network compared to a single IDS on a network. The artifact with more tuning would be able to handle a variety of DoS attacks without a drop in events going to the SIEM. As stated, there are many more test cases that can be tested to show more of the benefits of IDS being in a CIDS compared to a single IDS, there is a lot more work that can be done using this developed CIDS that will be discussed in Section 7 about future work.

## 7. FUTURE WORK AND RESEARCH DIRECTIONS

There are many areas for future work and research directions that can be built on this paper. In the following, we focus mainly on the integration of Artificial Intelligence (AI) and Machine Learning (ML) into the developed CIDS.

### 7.1 Revised AI-Assisted Architecture

Our immediate future work will focus on developing a decoupled hybrid approach for CIDS, which leverages the combined strengths of HIDS and NIDS. The idea being that these components operate independently and are tailored to individual nodes. Event feeds from these nodes are supplied to the central node and in return bespoke threat intelligence feeds are received. The central node will utilise AI to standardise and aggregate the events, prior to holistically analysing the data to identify threats and generating the bespoke threat intelligence feeds. The major benefit here is that by utilising this combined data, deeper patterns and temporal relationships can be identified between network and host behavioural characteristics surrounding triggered events.

### 7.2. Enhanced Threat Intelligence with ML

The classification of network activity as normal or malicious is a fundamental challenge in designing a CIDS. ML can identify patterns and behaviours indicative of security threats with high precision, given the availability of sufficiently large and detailed datasets that correctly encode characteristics of interest. In principle this is a typical classification problem, where the goal is to accurately assign labels (e.g. normal vs malicious) to relevant data. A great deal of work has already been undertaken in this area [20, 21, 22]. However, this seemingly straightforward classification task is complicated by the practical realities of dynamic network environments, human behaviour, and various interrelated factors at both the network and host levels, creating several challenges that impede ML which have not been fully accounted for. These include, for example,

- Nonstationarity: Evolving network activity, user behaviour, and the emergence of new threats result in concept drift [23, 24], where the data distributions describing threat and non-threat behaviours used to train ML algorithms shift over time. This undermines the assumptions of most static ML models, which typically rely on stationary data, necessitating adaptive approaches to maintain accuracy in changing environments.
- Class imbalance: This issue arises when one class is significantly rarer than another, either due to genuine scarcity or systematic biases in data collection [25, 26]. Many ML algorithms tend to favour the dominant, majority class, leading to poor performance on highly imbalanced datasets. In the context of CIDS, malicious activity often represents only a small fraction of observed traffic, making it challenging for ML models to effectively detect rare but critical events. Class imbalance also leads to increased false positive rates in practice, which can overwhelm analysts and reduce trust in the system.
- Temporal dependencies: Attack patterns often develop over time, making sequential relationships between events critical for accurate detection. Failure to account for these temporal dependencies can limit an ML model's ability to capture the context necessary for identifying evolving threats. Indeed, numerous approaches have been proposed to deal with such issues arising in general time-series data sets [27, 28].
- Novelty Emergence: Emerging threats, such as zero-day attacks, introduce patterns that are entirely absent from training data. This presents a unique challenge, as models must generalise beyond known behaviours and identify anomalies indicative of previously unseen attacks. A variety of approaches have been applied to this problem – one class learning [29], outlier detection [30], anomaly detection [31], and novel class detection [32].

The implementation of ML-based CIDS faces additional practical challenges, particularly in the context of a hybrid approach that combines HIDS and NIDS. High-dimensional, heterogeneous data from network and host sources must be standardised and aggregated for meaningful analysis. This process is computationally demanding, especially given the need for real-time processing to generate timely threat intelligence. Scalability is also essential in hybrid architectures, as event feeds from distributed nodes must be efficiently processed on a central server node. Lastly, interpretability remains critical, ensuring that the ML-supported threat intelligence feeds provide actionable and interpretable insights. Effectively addressing these challenges demands a holistic approach that combines multiple ML techniques to balance predictive power with operational constraints.

### 7.3. Public Dataset Creation

Existing datasets for IDS research often provide a narrow perspective, focusing on single systems or central observation points. This limitation is particularly evident in CIDS research, which aims to tackle complex modern networks. To address this, we aim to create a new public dataset that captures the nuances of modern network dynamics and architecture. The intention is that this holistic dataset will incorporate a diverse range of viewpoints within a network, facilitating more in-depth studies into how differing relationships, dynamics, and structures of modern networks impact on security collaboration and detection of cyber threats.

### 7.4. Conclusion

In this paper, we presented a CIDS leveraging Snort IDS nodes and centralised data analysis via the LogScale SIEM platform. The proposed system addresses the limitations of standalone IDS deployments by facilitating real-time data sharing, aggregation, and advanced threat analysis. By implementing Snort sensors across a simulated network, the study demonstrated that collaborative detection significantly enhances the ability to identify complex and distributed attacks, such as port scans and ICMP flood attacks, while reducing false positives through centralised correlation. The system evaluation highlighted its scalability, accuracy, and effectiveness in detecting and responding to threats. During testing, the CIDS efficiently processed logs from multiple nodes without significant delays. However, challenges such as alert fatigue caused by bottlenecks in rsyslog indicate areas for further refinement. Addressing these issues through optimised configurations and rule tuning could further improve performance and usability. This research underscores the potential of collaborative approaches to intrusion detection in modern network environments. It also opens avenues for integrating machine learning and artificial intelligence into the CIDS framework, enabling adaptive threat intelligence and improved anomaly detection against a broader range of attack scenarios. In conclusion, the proposed CIDS provides a robust foundation for improving intrusion detection through collaboration, paving the way for more adaptive and resilient intrusion detection solutions.